\documentclass [12 pt]{article}
\usepackage{psfig}
\usepackage{epsfig}
\newcommand {\Wb} {\bar{W}}
\newcommand {\Delb} {\bar{\Delta}}
\begin{document}
\title { Intrabeam scattering growth rates for a 
bi-gaussian distribution.}
\author{George Parzen}
\date{SEPTEMBER 2004 \\BNL REPORT  C-A/AP NO.169 }
\maketitle
\begin{abstract}
This note finds results for the intrabeam scattering growth rates 
for a bi-gaussian distribution.
The bi-gaussian distribution is interesting for studying the 
possibility of using electron cooling in RHIC. Experiments and computer studies 
indicate that in the presence of electron cooling, the beam distribution 
changes so that it developes a strong core and a long tail which is not described 
well by a gaussian, but may be better described by a bi-gaussian.
Being able to compute the effects of intrabeam scattering for a bi-gaussian
distribution would  be useful in computing the effects of electron cooling,
which depend critically on the details of the intrabeam scattering.
The calculation is done using the reformulation of intrabeam 
scattering theory given in [1] based on the treatments given 
by A. Piwinski [2] and J. Bjorken and S.K.
Mtingwa [3].
The bi-gaussian distribution is defined below as the sum of two gaussians in the 
particle coordinates $x,y,s,p_x,p_y,p_s$. The gaussian with the smaller dimensions
produces most of the core of the beam, and the gaussian with  the larger 
dimensions largely produces the long tail of the beam. The final 
result for the growth rates are expressed as the sum of three terms which can be
interperted respectively as the contribution to the growth rates due 
to the scattering of the particles in the first gaussian from themselves,
the scattering of the particles in the second gaussian from themselves, and
the scattering of the particles in the first gaussian from the 
particles in the second gaussian.
\end{abstract}

\section{Introduction}

This note finds results for the intrabeam scattering growth rates 
for a bi-gaussian distribution.

The bi-gaussian distribution is interesting for studying the 
possibility of using electron cooling in RHIC. Experiments and computer studies 
indicate that in the presence of electron cooling, the beam distribution 
changes so that it developes a strong core and a long tail which is not described 
well by a gaussian, but may be better described by a bi-gaussian.
Being able to compute the effects of intrabeam scattering for a bi-gaussian
distribution would  be useful in computing the effects of electron cooling,
which depend critically on the details of the intrabeam scattering.
The calculation is done using the reformulation of intrabeam scattering theory  
given in [1] based on the treatments given by 
A. Piwinski [2] and by J. Bjorken and S. 
Mtingwa [3].
The bi-gaussian distribution is defined below as the sum of two gaussians in the 
particle coordinates $x,y,s,p_x,p_y,p_s$. The gaussian with the smaller dimensions
produces most of the core of the beam, and the gaussian with  the larger 
dimensions largely produces the long tail of the beam. The final 
result for the growth rates are expressed as the sum of three terms which can be
interperted respectively as the contribution to the growth rates due 
to the scattering of the particles in the first gaussian from themselves,
the scattering of the particles in the second gaussian from themselves, and
the scattering of the particles in the first gaussian from the 
particles in the second gaussian.

\section{Basic results for intrabeam scattering}

This section lists some general results which can be used to find growth rates 
for a beam with any particle distribution $f(x.p)$. Following [3], 
growth rates will be computed for $<p_{i}p_{j}>$ , where the $<>$ indicates an 
average over all the particles in the bunch. From these one can compute the 
growth rates for the emittances, $<\epsilon_i>$. A result that holds in any 
coordinate system  and for any particle distribution $f(x.p)$ is given in [1]
as
\begin{eqnarray} 
\delta <(p_{i}p_{j}) > &=& N \int \;\: d^3x \frac {d^3p_1}{\gamma_1}
      \frac {d^3p_2}{\gamma_2} f(x,p_1)f(x,p_2) F(p_1,p_2)  C_{ij} dt \nonumber\\
C_{ij} &=& \pi \int_{0}^{\pi} d\theta \sigma(\theta) \sin^3 \theta \;
       \Delta^2 [\delta_{ij}-3\frac{\Delta_i \Delta_j}{\Delta^2} 
       +\frac{W_iW_j}{W^2}] \;\; i,j=1,3   \nonumber\\
\Delta_i &=& \frac{1}{2}(p_{1i}-p_{2i})   \nonumber\\
W_i &=& p_{1i}+p_{2i}
\end{eqnarray}

$Nf(x,p)$ gives the number of particles in $d^3xd^3p$, where N is the 
number of particles in a bunch. $\delta <(p_{i}p_{j}) >$ is the change in
$<(p_{i}p_{j}) >$ due to all particle collisions in the time interval dt.
The invariants $F(p_1,p_2),\Delta^2,W^2$ are given by
\begin{eqnarray*}
F(p_1,p_2) &=& c \frac {[(p_1p_2)^2-m_1^2m_2^2c^4]^{1/2}}{m_1m_2 c^2}  \nonumber\\
F(p_1,p_2) &=& \gamma_1 \gamma_2 c [(\vec{\beta_1}-\vec{\beta_2})^2 -(\vec{\beta_1} \times \vec{\beta_2})^2]^{1/2}                                \nonumber\\
\Delta^2 &=& \vec{\Delta}^2-\Delta_0^2, \;\:\;\:\Delta_0=(E_1-E_2)/(2c) \nonumber\\
W^2 &=& \vec{W}^2-W_0^2, \;\:\;\:W_0=(E_1+E_2)/c                     
\end{eqnarray*}

Eq.(1) is considerably simplified by going to the rest CS , which is the CS
moving along with the bunch and the particle motion is non-relativistic, and
putting $\sigma$ equal to the Coulomb cross section. One gets
\begin{eqnarray} 
\frac{1}{p_0^2}<\delta (p_{1i}p_{1j}) > &=& N \int \;\: d^3x d^3p_1
      d^3p_2 f(x,p_1)f(x,p_2) 2 \bar{\beta}c \; C_{ij}\;dt  
      \nonumber\\
\Delta_i &=& \frac{1}{2}(p_{1i}-p_{2i})   \nonumber\\
\bar{\beta}c &=& |\vec{\Delta}|/m         \nonumber\\
C_{ij} &=& \frac{2 \pi}{p_0^2} (r_0/2 \bar{\beta}^2)^2 
      \ln(1+(2 \bar{\beta}^2 b_{max}/r_0)^2) \;\; \nonumber\\ 
       & & [|\vec{\Delta}|^2 \delta_{ij}-3\Delta_i \Delta_j ] \;\; i,j=1,3     \nonumber\\
r_0 &=& Z^2e^2/mc^2                    \nonumber\\
\sigma(\theta) &=& [\frac{r_0}{2\bar{\beta}^2 }]^2 \frac{1}{(1-\cos \theta)^2} \nonumber\\
\cot (\theta_{min}/2) &=& 2 \bar{\beta}^2 b_{max}/r_0
\end{eqnarray}
$b_{max}$ is the largest allowed impact parameter in the center of mass CS. It has been asumed that
one can replace $\ln(1+(2 \bar{\beta}^2 b_{max}/r_0)-1$ by $\ln(1+(2 \bar{\beta}^2 b_{max}/r_0)$.

In Eq.(1), the original 11-dimensional integral which arises from 
intrabeam scattering theory has been reduced in [1] to a 9-dimensional integral by
integrating over all possible scattering angles. In [1] this reduction was 
done for any
particle distribution, $f(x,p)$. In [3], Bjorken and Mtingwa  first do the 
integration over $x,p_1,p_2$ using a simple gaussian distribution
before doing the integration over the scattering angles and no general result for
doing this reduction for any $f(x,p)$  is given.
In [2] Piwinski computes the growth rates for the emittances $<\epsilon_i>$ 
instead of for $<p_{i}p_{j}>$. A general result for reducing the integral by
integrating over all possible scattering angles, for any $f(x,p)$, for the
growth rates of $<\epsilon_i>$ is given. However, using this result for a 
complicated  distribution like the bi-gaussian would be difficult.

\section{Gaussian distribution}

We will first consider the case of a gaussian particle distribution. This will 
provide a more simple example of using the results in the reformulation given 
in [1]
and of the methods used to evaluate the integrals. Afterwards, the same procedures 
will be applied to the case of the bi-gaussian distribution.

Let $Nf(x,p)$ gives the number of particles in $d^3xd^3p$, where N is the 
number of particles in a bunch.
For a gaussian distribution, $f(x,p)$ ls given by
\begin{eqnarray} 
f(x,p)&=&\frac{1}{\Gamma} exp[-S(x,p)]  \nonumber\\
\Gamma &=& \int d^3xd^3p \;  exp[-S(x,p)] 
\end{eqnarray}
%
\begin{eqnarray}
S &=& S_x+S_y+S_s      \nonumber\\ 
  & &                  \nonumber\\                
S_x &=&\frac{1}{\bar{\epsilon_x}} \epsilon_x (x_\beta,x_{\beta}') \nonumber\\
x_\beta &=& x-D(p-p_0)/p_0     \nonumber\\
x_{\beta}' &=& x'-D'(p-p_0)/p_0 \;\;\; x'=p_x/p_0   \nonumber\\
\epsilon_x (x,x') &=& [x^2+(\beta_x x'+\alpha_x x)^2]/\beta_x \nonumber\\
  & &                  \nonumber\\ 
S_y &=& \frac{1}{\bar{\epsilon_y}}  \epsilon_y (y,y')\;\;\; y'=p_y/p_0  \nonumber\\
\epsilon_y (y,y') &=& [y^2+(\beta_y y'+\alpha_y y)^2]/\beta_y    \nonumber\\
  & &                  \nonumber\\ 
S_s &=& \frac{1}{\bar{\epsilon_s}}  \epsilon_s (s-s_c,(p-p_0)/p_0)  \nonumber\\
\epsilon_s (s-s_c,(p-p_0)/p_0)&=& \frac{(s-s_c)^2}{2\sigma_s^2}+\frac{((p-p_0)/p_0)^2}{2 \sigma_p^2} \nonumber\\
\epsilon_s (s-s_c,(p-p_0)/p_0)&=& \frac{1}{\beta_s} (s-s_c)^2+\beta_s ((p-p_0)/p_0)^2 \nonumber\\
\epsilon_s (s-s_c,(p-p_0)/p_0) &=& [(s-s_c)^2+(\beta_s ((p-p_0)/p_0))^2]/\beta_s \nonumber\\
\beta_s &=& \sigma_s/\sigma_p  \nonumber\\
\bar \epsilon_s &=& 2 \sigma_s \sigma_p  
\end{eqnarray}
$D$ is the horizontal dispersion. $D'=dD/ds$. 
A longitudinal emittance has been introduced
so that the longitudinal motion and the transverse motions can be treated in 
a similar manner. $s_c$ locates the center of the bunch. 

$\Gamma$ can now be 
computed using Eq.(1).This will provide an example how the integrals are done in 
this paper. The integration methods used here are somewhat more complicated
than those used in [3] but they will also work for the more complicated
bi-gaussian distribution.
\begin{eqnarray*} 
\Gamma &=& \int d^3xd^3p \;  exp[-S_x-S_y-S_s]
\end{eqnarray*}
Writing $\Gamma$ as $\Gamma=\Gamma_y \Gamma_{xs}$ and computing $\Gamma_y$
first because this part is simpler,
\begin{eqnarray} 
\Gamma_y &=&  \int dydp_y \;  exp[-S_y]         \nonumber\\
S_y &=& \frac{1}{\bar{\epsilon_y}}  \epsilon_y (y,y')  \;\;\; y'=p_y/p_0     \nonumber\\
\epsilon_y (y,y') &=& [y^2+2 +(\beta_y y'+\alpha_y y)^2]/\beta_y        \nonumber\\
\eta_y&=& y/\sqrt{\beta_y}, \;\;\; p_{\eta y}= (\beta_y y'+\alpha_y y)/\sqrt{\beta_y}
                                        \nonumber\\
dy dp_y &=& p_0 d\eta_y dp_{\eta y}                \nonumber\\
\Gamma_y &=&  p_0 \int d\eta_y dp_{\eta y} \; exp[-(\eta_y^2+p_{\eta y}^2)/\bar{\epsilon_y}]                                    \nonumber\\
\Gamma_y &=& \pi \bar{\epsilon_y} p_0       
\end{eqnarray}
Now for the remaining integral we have
\begin{eqnarray} 
\Gamma_{xs} &=&  \int dxdp_x dsdp_s\;  exp[-S_x-S_s]         \nonumber\\ 
\Gamma_{xs} &=&  \int dsdp_s\; exp[-S_s] \int dxdp_x \; exp[-S_x]    \nonumber\\
\mbox{Make the transformation} & &               \nonumber\\
x_{\beta} &=& x-D(p-p_0)/p_0          \nonumber\\
x'_{\beta }&=& x' -D'(p-p_0)/p_0    \nonumber\\
x' &=& p_x/p_0, \;\;\; x'_{\beta }=p_{\beta x}/p_0               \nonumber\\
     dxdp_x&=&p_0 dx_{\beta}dx_{\beta }'             \nonumber\\
  & &                                 \nonumber\\
\int dxdp_x \; exp[-S_x] &=& p_0 \int dx_{\beta}dx_{\beta}' \; exp[-S_x]   \nonumber\\
S_x &=&\frac{1}{\bar{\epsilon_x}} \epsilon_x (x_\beta,x_{\beta}') \nonumber\\
\int dxdp_x \; exp[-S_x] &=& \pi \bar{\epsilon_x} p_0 \mbox{   as in evaluating $\Gamma_y$ }          \nonumber\\
 & & \nonumber\\
\mbox{ $p\sim p_s$ in the Lab. CS and} & & \nonumber\\
\Gamma_{xs} &=& \pi^2 \bar{\epsilon_s}  \bar{\epsilon_x} p_0^2     \nonumber\\
  & &                                                \nonumber\\
\Gamma &=& \pi^3 \bar{\epsilon_s} \bar{\epsilon_x} \bar{\epsilon_y} p_0^3
\end{eqnarray}

\section{Growth rates for a Gaussian distribution}

In the following,the growth rates are given in the Rest Coordinate System,
which is the coordinate system moving along with the bunch.
Growth rates are given for $<p_i p_j>$. From these one can compute the 
growth rates for $<\epsilon_i>$. Using the general result, Eq.(2), one gets

\begin{eqnarray} 
\frac{1}{p_0^2}<\delta (p_{i}p_{j}) > &=& \frac {N}{\Gamma^2} \int \;\: d^3x d^3p_1
      d^3p_2 exp[-S(x,p_1)-S(x,p_2)] 2 \bar{\beta}c \; C_{ij}\;dt  
      \nonumber\\
\vec{\Delta} &=& \frac{1}{2}(\vec{p_{1}}-\vec{p_{2}})   \nonumber\\
\bar{\beta}c &=& |\vec{\Delta}|/m         \nonumber\\
C_{ij} &=& \frac{2 \pi}{p_0^2} (r_0/2 \bar{\beta}^2)^2 
      \ln(1+(2 \bar{\beta}^2 b_{max}/r_0)^2) \;\;  
       [|\vec{\Delta}|^2 \delta_{ij}-3\Delta_i \Delta_j
       ] \;\; i,j=1,3                             \nonumber\\
r_0 &=& Z^2e^2/mc^2                \nonumber\\
\Gamma &=& \pi^3 \bar{\epsilon_s} \bar{\epsilon_x} \bar{\epsilon_y} p_0^3
\end{eqnarray}

Transform to $W,\Delta$
\begin{eqnarray} 
p_1 &=& \frac{W}{2}+\Delta, \;\;\;\;\;\;\;\; p_2=\frac{W}{2}-\Delta  \nonumber\\
W &=& p_1+p_2, \;\;\;\;\;\;\;\:\;\;\;\;      \Delta=\frac {p_1-p_2}{2} \nonumber\\
d^3p_1 d^3p_2 &=& d^3Wd^3\Delta 
\end{eqnarray}

We will first do the integral over $d^3x$ and over $d^3W$.
For the y  part of the integral
\begin{eqnarray*}
S_y(y,p_{1y}) &=& \frac{1}{\bar{\epsilon_y}}  \epsilon_y (y,y_1'),\;\;\; y_1'=p_{1y}/p_0  \nonumber\\
\epsilon_y (y,y_1') &=& [y^2+(\beta_y y_1'+\alpha_y y)^2]/\beta_y    \nonumber\\
S_y(y,p_{1y}) &=& [y^2+(\beta_y (\frac{W_y}{2}+\Delta_y)/p_0 +\alpha_y y)^2]/(\beta_y \bar{\epsilon_y} )
\end{eqnarray*}
\begin{eqnarray} 
 S_y(y,p_{1y})+S_y(y,p_{2y})&=&(2y^2/\beta_y+2(\beta_y (W_y/p_0)/2+\alpha_y y)^2/\beta_y     \nonumber\\
  & & +2\beta_y^2 (\Delta_y/p_0)^2/\beta_y)) / \bar{\epsilon_y}   \nonumber\\
\mbox{Make the transformation } & &   \nonumber\\
\eta_y &=& \sqrt{2}y/\sqrt{\beta_y}, \;\;\;\;\;\; p_{\eta y}=\sqrt{2} (\beta_y(W_y/p_0)/2+\alpha_y y) /\sqrt{\beta_y}                       \nonumber\\
dydW_y &=& p_0 d\eta_y dp_{\eta y}
\end{eqnarray}
Integrate over $dy,dW_y$
\begin{eqnarray}
 \int  dy dW_y exp[-S_y(y,p_{1y})-S_y(y,p_{2y})]&=& p_0 \int  d\eta_y dp_{\eta y} 
\nonumber\\
 & & exp[-\frac {\eta_y^2+p_{\eta y}^2+2\beta_y^2 (\Delta_y/p_0)^2/\beta_y } {\bar{\epsilon_y}}]         \nonumber\\
  &=& p_0 \pi \bar{\epsilon_y} exp[-\frac{2 \beta_y}{\bar{\epsilon_y}}
(\Delta_y/p_0)^2]                              \nonumber\\
  &=& p_0 \pi \bar{\epsilon_y} sxp[-R_y]              \nonumber\\
 & &                              \nonumber\\
R_y &=& \frac{2 \beta_y}{\bar{\epsilon_y}}(\Delta_y/p_0)^2
\end{eqnarray}

In doing the remainder of the integral, the integral over $dxdW_xdsdW_s$ we will
do the integral over $dxdW_x$ first and then the integral over $dsdW_s$.  Note 
that the integral is being done in the Rest CS and in the expression for $S_x$
one has to replace $p-p_0 \sim p_s-p_0$ in the Lab CS by $\gamma p_s$ in the Rest
CS. Remember also that $f(x,p)$ is an invariant (see [1])
One finds for $S_x(x,p_{1x})$ 
\begin{eqnarray}
S_x(x,p_{1x}) &=& \{ [x-\gamma D\bar{W}_s/2-\gamma D \bar{\Delta}_s]^2+
    [ \beta_x (\bar{W}_x/2+\bar{\Delta}_x-\gamma D'\bar{W}_s/2-\gamma D'  
    \bar{\Delta}_s) +                                \nonumber\\
 & & \alpha_x (x-\gamma D\bar{W}_s/2-\gamma D \bar{\Delta}_s) ]^2 \} /
    (\beta_x \bar{\epsilon_x} )  \nonumber\\
 & &                  \nonumber\\
\bar{W}_i &=& W_i/p_0 \;\;\;\;\;\;\bar{\Delta}_i=\Delta/p_0    \nonumber\\
 & &                  \nonumber\\
S_x(y,p_{1x}) &=& \{ [x-\gamma D\bar{W}_s/2-\gamma D \bar{\Delta}_s]^2+
[ \beta_x (\bar{W}_x/2- \gamma D'\bar{W}_s/2) +\nonumber\\
 & & \alpha_x (x-\gamma D\bar{W}_s/2)+
(\beta_x\bar{\Delta}_x-\gamma \bar{D} \bar{\Delta}_s) ]^2   \} /
(\beta_x \bar{\epsilon_x} )  \nonumber\\
\bar{D} &=& \beta_x D'+\alpha_x D
\end{eqnarray}
we then find for $S_x(x,p_{1x})+S_x(x,p_{2x})$
\begin{eqnarray}
S_x(x,p_{1x})+S_x(x,p_{2x}) &=& \{ 2[x-\gamma D\bar{W}_s/2]^2+2 \gamma^2 D^2\bar{\Delta}_s^2+   \nonumber\\
 & &   2[ \beta_x (\bar{W}_x/2-\gamma D'\bar{W}_s/2) + 
     \alpha_x (x-\gamma D\bar{W}_s/2) ]^2+ \nonumber\\
 & &  2[\beta_x\bar{\Delta}_x-\gamma \bar{D} \bar{\Delta}_s ]^2 \} /
  (\beta_x \bar{\epsilon_x} )
\end{eqnarray}
Now make the transformations
\begin{eqnarray}
x^* &=& \sqrt{2}x-\gamma D \Wb_s/\sqrt{2} \;\;\;\;\;\; p_x^*=\Wb_x/\sqrt{2}-\gamma D'\Wb_s/\sqrt{2}                \nonumber\\
\eta_x &=& x^*/\sqrt{\beta_x}\;\;\;\;\;p_{\eta_x x}=(\beta_x p_x^*+\alpha_x x^*)/
   \sqrt{\beta_x}  \nonumber\\
dxdW_x &=& p_0 dx^*dp_x^*=p_0 d\eta_x dp_{\eta_x x}
\end{eqnarray}
Doing the integral over $dxdW_x$ one finds
\begin{eqnarray}
 \int  dx dW_x exp[-S_x(x,p_{1x})-S_x(x,p_{2x})]&=& 
      p_0 \int  d\eta_x dp_{\eta_x x} \nonumber\\
   & & exp[- \{ \eta_x^2+p_{\eta_x}^2+    \nonumber\\
    & &   2 [\gamma^2 D^2 \bar{\Delta}_s^2+(\beta_x \bar{\Delta}_x-\gamma \bar{D} \bar{\Delta}_s)^2] 
          / \beta_x )\} / \bar{\epsilon_x} ]         \nonumber\\
  &=& p_0 \pi \bar{\epsilon_x} exp[-R_x]                              \nonumber\\
 & &                                \nonumber\\
R_x &=&  2 [\gamma^2 D^2\bar{\Delta}_s^2+(\beta_x\bar{\Delta}_x-\gamma \bar{D} \bar{\Delta}_s)^2] 
       /  (\beta_x \bar{\epsilon_x} )           \nonumber\\
\end{eqnarray}

Now do the integral over $dsdW_s$. One may note that the form of the intgral here 
is similar to the integral done over $dydW_y$. The result is then the same with
the proper sustitutions of $s$ for $y$.
\begin{eqnarray}
 \int  dx dW_s exp[-S_s(s,p_{1s})-S_s(s,p_{2s})]&=& p_0 \pi \bar{\epsilon}_s 
                             exp[-R_s]              \nonumber\\
 & &                       \nonumber\\
R_s &=& \frac{2 \gamma^2 \beta_s}{\bar{\epsilon_s}}(\Delta_s/p_0)^2
\end{eqnarray}
Note that the term $\beta_s ((p-p_0)/p_0)^2$ in $S_s$ in the Lab. CS 
has to be replaced by $\gamma^2 \beta_s (p_s/p_0)^2$ in the Rest CS.

Using Eq.(7), one gets the result for  the growth rates in the Rest CS 
for a gauusian distribution.
\begin{eqnarray}
\frac{1}{p_0^2} \frac{d} {dt}<p_i p_j>  &=& \frac{N}{\Gamma} \int d^3\Delta \;
exp[-R] C_{ij}   \nonumber\\
C_{ij}&=& \frac{2 \pi}{p_0^2} (r_0/2\bar{\beta}^2)^2
        (|\Delta|^2 \delta_{ij}-3 \Delta_i \Delta_j ) 2\bar{\beta}c  \;\;
        ln[1+(2\bar{\beta}^2 b_{max}/r_0)^2]   \nonumber\\
\bar{\beta} &=& \beta_0 \gamma_0|\Delta/p_0| \nonumber\\
& &                                    \nonumber\\
r_0 &=& Z^2e^2/Mc^2    \nonumber\\
\Gamma &=& \pi^3 \bar{\epsilon_s} \bar{\epsilon_x} \bar{\epsilon_y} p_0^3  \nonumber\\
 & &                             \nonumber\\
R &=& R_x+R_y+R_s  \nonumber\\
R_x &=& \frac{2}{\beta_x \bar{\epsilon_x}} [\gamma^2 D^2 \Delta_s^2 +
        (\beta_x \Delta_x-\gamma \tilde{D} \Delta_s)^2 ]/p_0^2  \nonumber\\
   \tilde{D} &=& \beta_x D'+\alpha_x D   \nonumber\\
R_y &=& \frac{2\beta_y}{ \bar{\epsilon_y}}  \Delta_y^2/p_0^2 \nonumber\\
R_s &=& \frac{2\beta_s}{ \bar{\epsilon_s}} \gamma^2 \Delta_s^2/p_0^2  
\end{eqnarray}

The integral over $d^3 \Delta$ is an integral over all possible values 
of the relative momemtum for any two particles in a bunch. $\beta_0,\gamma_0$
are the beta and gamma corresponding to $p_0$, the central momemtum of the
bunch in the Laboratory Coordinate System. $\gamma=\gamma_0$

The above 3-dimensional integral can be reduced to a 2-dimensional integral
by integrating over $|\Delta|$ and using $d^3\Delta=|\Delta|^2 d|\Delta|
 sin\theta d\theta d\phi$. This gives
\begin{eqnarray}
\frac{1}{p_0^2} \frac{d} {dt}<p_i p_j>  &=& \frac{N}{\Gamma} 
   2 \pi p_0^3 \left( \frac{r_0}{2\gamma_0^2 \beta_0^2}\right)^2 2\beta_0 \gamma_0c
      \int sin\theta d\theta d\phi \; (\delta_{ij}-3 g_ig_j)   \nonumber\\   
  & &      \frac{1}{F} ln\left[\frac{\hat{C}}{F}\right]   \nonumber\\
    g_3&=&cos\theta=g_s  \nonumber\\
    g_1&=&sin\theta cos\phi=g_x   \nonumber\\
    g_2&=&sin\theta sin\phi=g_y   \nonumber\\
    \hat{C}&=&2 \gamma_0^2 \beta_0^2 b_{max}/r_0   \nonumber\\
 & &                     \nonumber\\
    F&=&R/(|\Delta|/p_0)^2   \nonumber\\
    F &=& F_x+F_y+F_s  \nonumber\\
F_x &=& \frac{2}{\beta_x \bar{\epsilon_x}} [\gamma^2 D^2 g_s^2 +
        (\beta_x g_x-\gamma \bar{D} g_s)^2 ]  \nonumber\\
F_y &=& \frac{2}{ \bar{\epsilon_y}} \beta_y g_y^2  \nonumber\\
F_s &=& \frac{2}{\bar{\epsilon_s}} \beta_s \gamma^2 g_s^2 
\end{eqnarray}
In obtaining the above, one uses $z=|\Delb |^2,dz=2|\Delb | d|\Delb |$ and
\[\int_ {0}^{\infty} dz \;\; exp[-Fz] ln[\hat{C}z]=\frac {1}{F}[ln\left[\frac{\hat{C}}{F}\right]-.5772]  \]
For $Z=80,A=200,\gamma=100,b_{max}=1 cm$, $\log_{10} \hat{C}=18.6$

\section{Bi-Gaussian distribution}
The bi-gaussian distribution will be assumed to have the form given 
by the following. 

$Nf(x,p)$ gives the number of particles in $d^3xd^3p$, where N is the 
number of particles in a bunch.
For a bi-gaussian distribution, $f(x,p)$ ls given by
\begin{eqnarray} 
f(x,p) &=& \frac{N_a}{N}\frac{1}{\Gamma_a} exp[-S_a(x,p)]+
      \frac{N_b}{N}\frac{1}{\Gamma_b} exp[-S_b(x,p)]              \nonumber\\
\Gamma_a &=& \pi^3 \bar{\epsilon_{sa}} \bar{\epsilon_{xa}} \bar{\epsilon_{ya}} p_0^3  \nonumber\\
\Gamma_b &=& \pi^3 \bar{\epsilon_{sb}} \bar{\epsilon_{xb}} \bar{\epsilon_{yb}} p_0^3  
\end{eqnarray}
In the first gaussian,to find $\Gamma_a,S_a$ then in the expressions 
for $\Gamma,S$, given above for the gaussian distribution, replace 
$\bar{\epsilon_x},\bar{\epsilon_y},\bar{\epsilon_s}$ by $\bar{\epsilon_{xa}},
\bar{\epsilon_{ya}},\bar{\epsilon_{sa}}$.
In the second gaussian, in the expressions for $\Gamma,S$, replace 
$\bar{\epsilon_x},\bar{\epsilon_y},\bar{\epsilon_s}$ by $\bar{\epsilon_{xb}},
\bar{\epsilon_{yb}},\bar{\epsilon_{sb}}$.
In addition. $N_a+N_b=N$. This bi-gaussian has 7 parameters instead of 
the three parameters of a gaussian.

\section{Growth rates for a Bi- Gaussian distribution}

In the following,the growth rates are given in the Rest Coordinate System,
which is the coordinate system moving along with the bunch.
Growth rates are given for $<p_i p_j>$. From these one can compute the 
growth rates for $<\epsilon_i>$.Starting with Eq.2 and using the $f(x,p)$
from Eq.18, one gets
\begin{eqnarray}
 & &              \nonumber\\        
\frac{1}{p_0^2}<\delta (p_{i}p_{j}) > &=&  \int \;\: d^3x d^3p_1 d^3p_2 
      \left [ \frac{N_a}{N}\frac{1}{\Gamma_a} exp[-S_a(x,p_1)]+
          \frac{N_b}{N}\frac{1}{\Gamma_b} exp[-S_b(x,p_1)] \right ]   \nonumber\\
 & &      \left [ \frac{N_a}{N}\frac{1}{\Gamma_a} exp[-S_a(x,p_2)]+
          \frac{N_b}{N}\frac{1}{\Gamma_b} exp[-S_b(x,p_2)] \right ] \nonumber\\   
 & &      2 \bar{\beta}c \; C_{ij}\;dt  
      \nonumber\\
\vec{\Delta} &=& \frac{1}{2}(\vec{p_{1}}-\vec{p_{2}})   \nonumber\\
\bar{\beta}c &=& |\vec{\Delta}|/m         \nonumber\\
C_{ij} &=& \frac{2 \pi}{p_0^2} (r_0/2 \bar{\beta}^2)^2 
      \ln(1+(2 \bar{\beta}^2 b_{max}/r_0)^2) \;\;  
       [|\vec{\Delta}|^2 \delta_{ij}-3\Delta_i \Delta_j
       ] \;\; i,j=1,3                             \nonumber\\
r_0 &=& Z^2e^2/mc^2
\end{eqnarray}
The term in the integrand which contains $exp[-S_a(x,p_1)-S_a(x,p_2)]$ is similar
to the integrand for the gaussian distribution except that $\bar{\epsilon}_{i}$
are replaced by $\bar{\epsilon}_{ia}$ and leads to the same result as that 
given by Eq.(16) for the gaussian beam except that $R$ has to be replaced by $R_a$
where $R_a$ is obtained from $R$ by replacing $\bar{\epsilon}_{i}$ by 
$\bar{\epsilon}_{ia}$. The term containing $exp[-S_b(x,p_1)-S_b(x,p_2)]$ can be 
evaluated in the same way leading to the same result as that 
given by Eq.(16) for the gaussian beam except that $R$ has to be replaced by $R_b$
where $R_b$ is obtained from $R$ by replacing $\bar{\epsilon}_{i}$ by 
$\bar{\epsilon}_{ib}$. The only terms that need further evaluation are the the two 
cross product terms. The two cross product terms are equal because of the symmetry
of $p_1$ and $p_2$ in the rest of the integrand. This leads to the remaining 
integral to be evaluated
\[ \int \;\: d^3x d^3p_1 d^3p_2 
    \frac{2N_aN_b}{N^2}\frac{1}{\Gamma_a \Gamma_b} 
      exp[-S_a(x,p_1)-S_b(x,p_2)] \; 2 \bar{\beta}c \; C_{ij} \]

In evaluating this integral, we will use the same procedure as was used for the 
gaussian distribution. We will first transform to $W,\Delta$ from $p_1,p_2$
(see Eq.(8). We will then do the integral over $d^3x$ and over $d^3W$.
For the y  part of the integral one finds ,
\begin{eqnarray}
S_{ya}(y,p_{1y}) &=& \{ y^2+[\beta_y (\Wb_y/2+\Delb_y) +\alpha_y y ]^2 \}    
/(\beta_y \bar{\epsilon}_{ya} )   \nonumber\\
\Wb_y &=& W_y /p_0 \;\;\;\;\;\;   \Delb_y =\Delta_y /p_0
\end{eqnarray}
One then finds that
\begin{eqnarray}
  & &                                  \nonumber\\
  & &                                  \nonumber\\
 S_{ya}(y,p_{1y})+S_{yb}(y,p_{2y})&=&\{ 2y^2/\beta_y +2[\beta_y (\Wb_y/2+\alpha_y y) ]^2/\beta_y     \nonumber\\
  & & +2\beta_y^2 \Delb_y^2/\beta_y) \} / \bar{\epsilon}_{yc}+     \nonumber\\
  & & \{ 4 (\beta_y \Delb_y) (\beta_y \Wb_y/2+\alpha_y y)/\beta_y \} / \bar{\epsilon}_{yd}                                  \nonumber\\
\frac{1}{\bar{\epsilon}_{yc}} &=& \frac{1}{2} (\frac{1}{\bar{\epsilon}_{ya}}+
        \frac{1}{\bar{\epsilon}_{yb} } )           \nonumber\\
\frac{1}{\bar{\epsilon}_{yd}} &=& \frac{1}{2} (\frac{1}{\bar{\epsilon}_{ya}}-
        \frac{1}{\bar{\epsilon}_{yb}} )              \nonumber\\
 & &
\end{eqnarray}                                  
Make the transformation
\begin{eqnarray}
\eta_y &=& \sqrt{2}y/\sqrt{\beta_y}, \;\;\;\;\;\; p_{\eta y}=\sqrt{2} (\beta_y \Wb_y/2+\alpha_y y) /\sqrt{\beta_y}                       \nonumber\\
dydW_y &=& p_0 d\eta_y dp_{\eta y}
\end{eqnarray}
Integrate over $dy,dW_y$
\begin{eqnarray}
 \int  dy dW_y ecp[-S_{ya}(y,p_{1y})-S_{yb}(y,p_{2y})] &=& p_0 \int  d\eta_y dp_{\eta y} 
\nonumber\\
 & & exp[-\frac {\eta_y^2+p_{\eta y}^2+2\beta_y^2 (\Delta_y/p_0)^2/\beta_y } {\bar{\epsilon}_{yc}}+         \nonumber\\
  & & 4 \beta_y \Delb_y \frac{p_{\eta y}/(\sqrt{2} \sqrt{\beta_y})} {  \bar{\epsilon}_{yd} }]                                               \nonumber\\
  &=& p_0 \pi \bar{\epsilon_{yc}} exp[-\frac{2 \beta_y}{\bar{\epsilon}_{yc}}
\Delb_y^2 +\frac{2 \beta_y} {\bar{\epsilon}_{yd}^2/\bar{\epsilon}_{yc}} \Delb_y^2]                              \nonumber\\
  &=& p_0 \pi \bar{\epsilon}_{yc} exp[-R_{yc}+R_{yd}]              \nonumber\\
 & &                           \nonumber\\
R_{yc} &=& \frac{2 \beta_y} {\bar{\epsilon}_{yc}}\Delb_y^2  \nonumber\\
R_{yd} &=& \frac{2 \beta_y} {\bar{\epsilon}_{yd}^2/\bar{\epsilon}_{yc}} \Delb_y^2
\end{eqnarray}
The exponent $R_{yc}-R_{yd}$ has to be positive. This can be made more obvious by noting that
\[ \frac {1}{\bar{\epsilon}_{yc}^2}-\frac{1}{\bar{\epsilon}_{yd}^2}= \frac{1}{\bar{\epsilon}_{ya}\bar{\epsilon}_{yb}} \]

In doing the remainder of the integral, the integral over $dxdW_xdsdW_s$ we will
do the integral over $dxdW_x$ first and then the integral over $dsdW_s$.  Note 
that the integral is being done in the Rest CS and in the expression for $S_x$
one has to replace $p-p_0 \sim p_s-p_0$ in the Lab. CS by $\gamma p_s$ in the Rest
CS. Remember also that $f(x,p)$ is an invariant (see [1])
One finds for $S_{xa}(x,p_{1x})$ 
\begin{eqnarray}
S_{xa}(x,p_{1x}) &=& \{ [x-\gamma D\bar{W}_s/2-\gamma D \bar{\Delta}_s]^2+
    [\beta_x (\bar{W}_x/2+\bar{\Delta}_x-\gamma D'\bar{W}_s/2-\gamma D'  
    \bar{\Delta}_s) +                                \nonumber\\
 & & \alpha_x (x-\gamma D\bar{W}_s/2-\gamma D \bar{\Delta}_s) ]^2 \}/
    (\beta_x \bar{\epsilon_xa} )  \nonumber\\
 & &                  \nonumber\\
\bar{W}_i &=& W_i/p_0 \;\;\;\;\;\;\bar{\Delta}_i=\Delta/p_0    \nonumber\\
 & &                  \nonumber\\
S_{xa}(y,p_{1x}) &=& \{ [x-\gamma D\bar{W}_s/2-\gamma D \bar{\Delta}_s]^2+
[ \beta_x (\bar{W}_x/2- \gamma D'\bar{W}_s/2) +\nonumber\\
 & & \alpha_x (x-\gamma D\bar{W}_s/2)+
(\beta_x\bar{\Delta}_x-\gamma \bar{D} \bar{\Delta}_s) ]^2   \} /
(\beta_x \bar{\epsilon_{xa}} )  \nonumber\\
\bar{D} &=& \beta_x D'+\alpha_x D
\end{eqnarray}
we then find for $S_{xa}(x,p_{1x})+S_{xb}(x,p_{2x})$
\begin{eqnarray}
S_{xa}(x,p_{1x})+S_{xb}(x,p_{2x}) &=& \{2[x-\gamma D\bar{W}_s/2]^2+2 \gamma^2 D^2\bar{\Delta}_s^2+   \nonumber\\
 & &   2[ \beta_x (\bar{W}_x/2-\gamma D'\bar{W}_s/2) + 
     \alpha_x (x-\gamma D\bar{W}_s/2) ]^2+ \nonumber\\
 & &  2[\beta_x\bar{\Delta}_x-\gamma \bar{D} \bar{\Delta}_s]^2 \} /
  (\beta_x \bar{\epsilon}_{xc} ) +         \nonumber\\
 & & \{ -4 \gamma D \bar{\Delta}_s [x-\gamma D\bar{W}_s/2]/\beta_x  \nonumber\\
 & & +4 (\beta_x\bar{\Delta}_x-\gamma \bar{D} \bar{\Delta}_s) [ \beta_x (\bar{W}_x/2- \gamma D'\bar{W}_s/2) +         \nonumber\\
 & & \alpha_x (x-\gamma D\bar{W}_s/2) ] /\beta_x \} /\bar{\epsilon}_{xd}  \nonumber\\
  & &                         
\end{eqnarray}
Now make the transformations
\begin{eqnarray}
x^* &=& \sqrt{2}x-\gamma D \bar{W}_s/\sqrt{2} \;\;\;\;\;\; p_x^*=\bar{W}_x/\sqrt{2}-\gamma D'\bar{W}_s/\sqrt{2}                \nonumber\\
\eta_x &=& x^*/\sqrt{\beta_x}\;\;\;\;\;p_{\eta_x x}=(\beta_x p_x^*+\alpha_x x^*)/
   \sqrt{\beta_x}  \nonumber\\
dxdW_x &=& p_0 dx^*dp_x^*=p_0 d\eta_x dp_{\eta_x x}
\end{eqnarray}
Doing the integral over $dxdW_x$ one finds
\begin{eqnarray}
 \int  dx dW_x exp[-S_{xa}(x,p_{1x})-S_{xb}(x,p_{2x})]&=& 
      p_0 \int  d\eta_x dp_{\eta_x x} \nonumber\\
   & & exp[-\{ \eta_x^2+p_{\eta_x}^2    \nonumber\\
  & & +2[\gamma^2 D^2\bar{\Delta}_s^2+(\beta_x\bar{\Delta}_x-\gamma \bar{D} \bar{\Delta}_s)^2 ]/\beta_x \} /
     \bar{\epsilon}_{xc}        \nonumber\\
 & &                   \nonumber\\
 & & +\{-4 \gamma D \Delb_s \eta_x/\sqrt{2 \beta_x}        \nonumber\\
 & & +4 (\beta_x\bar{\Delta}_x-\gamma \bar{D} \bar{\Delta}_s) p_{\eta_x x}/\sqrt{2 \beta_x} \} /\bar{\epsilon}_{xd} ]        \nonumber\\
 & &                   \nonumber\\
  &=& p_0 \pi \bar{\epsilon_{xc}} exp[-R_{xc}+R_{xd}]                              \nonumber\\
 & &                            \nonumber\\
R_{xc} &=&  2[\gamma^2 D^2\bar{\Delta}_s^2+(\beta_x\bar{\Delta}_x-\gamma \bar{D} \bar{\Delta}_s)^2 ] /  (\beta_x \bar{\epsilon}_{xc} )        \nonumber\\
R_{xd} &=& 
        2 \{ [-\gamma D \Delb_s ]^2       \nonumber\\
 & & +[(\beta_x\bar{\Delta}_x-\gamma \bar{D} \bar{\Delta}_s) ]^2 \}  \nonumber\\
 & &  /(\beta_x \bar{\epsilon}_{xd}^2/ \bar{\epsilon}_{xc}) \nonumber\\
 & &
\end{eqnarray}

Now do the integral over $dsdW_s$. One may note that the form of the intgral here 
is similar to the integral done over $dydW_y$. The result is then the same with
the proper sustitutions of $s$ for $y$.
\begin{eqnarray}
 \int  dx dW_s exp[-S_{sa}(s,p_{1s})-S_{sb}(s,p_{2s})]&=& p_0 \pi 
    \bar{\epsilon}_{sc} exp[-R_{sc}+R_{sd}]              \nonumber\\
 & &              \nonumber\\
R_{sc} &=& \frac{2 \beta_s} {\bar{\epsilon}_{sc}}\Delb_s^2  \nonumber\\
R_{sd} &=& \frac{2 \beta_s} {\bar{\epsilon}_{sd}^2/\bar{\epsilon}_{sc}} \Delb_s^2
\end{eqnarray}
Note that the term $\beta_s ((p-p_0)/p_0)^2$ in $S_s$ in the Lab. CS 
has to be replaced by $\gamma^2 \beta_s (p_s/p_0)^2$ in the Rest CS.

Putting all the above results, for the bi-gaussian distribution, together one gets the final result
\begin{eqnarray}
\frac{1}{p_0^2} \frac{d} {dt}<p_i p_j>  &=& N \int d^3\Delta \;
    C_{ij} [ \left(\frac{N_a}{N}\right)^2 \frac{exp(-R_a)}{\Gamma_a}+
    \left(\frac{N_b}{N}\right)^2 \frac{exp(-R_b)}{\Gamma_b}   \nonumber\\
 & & +2\frac{N_a N_b}{N^2} \frac{\Gamma_c}{\Gamma_a \Gamma_b} exp(-T) ] \nonumber\\
 & &                          \nonumber\\        
 C_{ij}&=& \frac{2 \pi}{p_0^2} (r_0/2\bar{\beta}^2)^2
        (|\Delta|^2 \delta_{ij}-3 \Delta_i \Delta_j ) 2\bar{\beta}c  \;\;
        ln[1+(2\bar{\beta}^2 b_{max}/r_0)^2]   \nonumber\\
\bar{\beta} &=& \beta_0 \gamma_0|\Delta/p_0| \nonumber\\
r_0 &=& Z^2e^2/Mc^2    \nonumber\\
 & &                                     \nonumber\\
\frac{1}{\bar{\epsilon_{ic}}}&=& \frac{1}{2}\left(\frac{1}{\bar{\epsilon_{ia}}}+
     \frac{1}{\bar{\epsilon_{ib}}} \right) \;\;i=x,y,s   \nonumber\\
\frac{1}{\bar{\epsilon}_{id}} &=& \frac{1}{2} (\frac{1}{\bar{\epsilon}_{ia}}-
        \frac{1}{\bar{\epsilon}_{ib}} )              \nonumber\\
& &                                    \nonumber\\
r_0 &=& Z^2e^2/Mc^2    \nonumber\\
\Gamma_a &=& \pi^3 \bar{\epsilon_{sa}} \bar{\epsilon_{xa}} \bar{\epsilon_{ya}} p_0^3  \nonumber\\
 & &                             \nonumber\\
R_a &=& R_{xa}+R_{ya}+R_{sa}  \nonumber\\
R_{xa} &=& \frac{2}{\beta_x \bar{\epsilon_{xa}}} [\gamma^2 D^2 \Delta_s^2 +
        (\beta_x \Delta_x-\gamma \tilde{D} \Delta_s)^2 ]/p_0^2  \nonumber\\
   \tilde{D} &=& \beta_x D'+\alpha_x D   \nonumber\\
R_{ya} &=& \frac{2}{\beta_y \bar{\epsilon_{ya}}} \beta_y^2 \Delta_y^2/p_0^2 \nonumber\\
R_{sa} &=& \frac{2}{\beta_s \bar{\epsilon_{sa}}} \beta_s^2 \gamma^2 \Delta_s^2/p_0^2   \nonumber\\
 & &                                     \nonumber\\
T&=& T_x+T_y+T_s   \nonumber\\
T_x&=& R_{xc}- R_{xd}   \nonumber\\
T_y&=& R_{yc}- R_{yd}   \nonumber\\
T_s&=& R_{sc}-R_{sd}    \nonumber\\
 & &                                     \nonumber\\ 
 & &                                     \nonumber\\ 
R_{xd} &=& 
        2 \{ [-\gamma D \Delb_s  ]^2      \nonumber\\
 & & +[(\beta_x\bar{\Delta}_x-\gamma \bar{D} \bar{\Delta}_s) ]^2 \}  \nonumber\\
 & &  /(\beta_x \bar{\epsilon}_{xd}^2/ \bar{\epsilon}_{xc}) \nonumber\\
R_{yd} &=& \frac{2 \beta_y} {\bar{\epsilon}_{yd}^2/\bar{\epsilon}_{yc}} \Delb_y^2    \nonumber\\
R_{sd} &=& \frac{2 \beta_s} {\bar{\epsilon}_{sd}^2/\bar{\epsilon}_{sc}} \Delb_s^2 \nonumber\\
\Delb_i &=& \Delta_i/p_0
\end{eqnarray}

$R_a,R_b,R_c$ are each the same as $R_a$ given above  
except that $\bar{\epsilon_{ia}}$ are replaced by 
$\bar{\epsilon_{ia}},\bar{\epsilon_{ib}},\bar{\epsilon_{ic}}$ respectively.
The same remarks apply to $\Gamma_a,\Gamma_b,\Gamma_c$

The above 3-dimensional integral can be reduced to a 2-dimensional integral
by integrating over $|\Delta|$ and using $d^3\Delta=|\Delta|^2 d|\Delta|
 sin\theta d\theta d\phi$. This gives

\begin{eqnarray}
\frac{1}{p_0^2} \frac{d} {dt}<p_i p_j>  &=& 2 \pi p_0^3 \left( \frac{r_0}
    {2\gamma_0^2 \beta_0^2}\right)^2 2\beta_0 \gamma_0c
      \int sin\theta d\theta d\phi \; (\delta_{ij}-3 g_ig_j)  \nonumber\\
 & &N [ \left(\frac{N_a}{N}\right)^2 \frac{1}{\Gamma_a F_a} ln[\frac{\hat{C}}{F_a}]
  +\left(\frac{N_b}{N}\right)^2 \frac{1}{\Gamma_b F_b} ln[\frac{\hat{C}}{F_b}] \nonumber\\
 & &         +2 \frac{N_aN_b}{N^2} \frac{\Gamma_c}{\Gamma_a\Gamma_b}
           \frac{1}{G} ln[\frac{\hat{C}}{G} ] ]          \nonumber\\
 & &                         \nonumber\\
    g_3&=&cos\theta=g_s  \nonumber\\
    g_1&=&sin\theta cos\phi=g_x   \nonumber\\
    g_2&=&sin\theta sin\phi=g_y   \nonumber\\
    \hat{C}&=&2 \gamma_0^2 \beta_0^2 b_{max}/r_0   \nonumber\\
 & &                         \nonumber\\
F_i &=& R_i/(|\Delta|/p_0)^2  \;\;\;\;i=a,b,c      \nonumber\\   
G&=&T/(|\Delta|/p_0)^2  \nonumber\\
\end{eqnarray}

$F_a,F_b,F_c$ are each the same F that was defined for the Gaussian 
distribution except that the $\bar{\epsilon_{i}}$ are replaced by 
$\bar{\epsilon_{ia}},\bar{\epsilon_{ib}},\bar{\epsilon_{ic}}$ respectively.

The above results for the growth rates for a bi-gaussian distribution
are expressed as an integral which contains 3 terms, each of which is similar
to the one term in the results for the gaussian distribution. 
These three terms may be given a simple interpertation. The first term represents
the contribution to the growth rates due to the scattering of the $N_a$ particles
of the first gaussian from themselves, the seond term the contribution
due to the scattering of the $N_b$ particles of the
second gaussian from themselves, 
and the third term the contribution due to the scattering of the $N_a$ particles
of the first gaussian from the $N_b$ partcles of the second gaussian.

\section{Emittance growth rates}

One can compute growth rates for the average emittances, $<\epsilon_i>$ in the 
Laboratory Coordinate System, from the growth rates for $<p_ip_j>$ in the 
Rest Coordinate System.In the following , $dt$ is the time interval in the
Laboratory System and $d\tilde{t}$ is the time interval in the Rest System.
$dt=\gamma d\tilde{t}$

\begin{eqnarray}
\frac{d}{dt} <\epsilon_x> &=& \frac{\beta_x}{\gamma} 
        \frac{d}{d\tilde{t}} <p_x^2/p_0^2> + \frac{D^2+\tilde{D}^2}{\beta_x} \gamma
         \frac{d}{d\tilde{t}}<p_s^2/p_0^2> 
        -2 \tilde{D} \frac{d}{d\tilde{t}}<p_x p_s/p_0^2>     \nonumber\\
\frac{d}{dt} <\epsilon_y> &=& \frac{\beta_y}{\gamma} \frac{d}{d\tilde{t}}
        <p_y^2/p_0^2>      \nonumber\\
\frac{d}{dt} <\epsilon_s> &=& \beta_s \gamma \frac{d}{d\tilde{t}}
        <p_s^2/p_0^2>      
\end{eqnarray}

To derive the above results, the simplest case to treat is that of the vertical 
emittance. The verical emmitance is given by
\begin{eqnarray}
\epsilon_y (y,y') &=& [y^2+(\beta_y y'+\alpha_y y)^2]/\beta_y    \nonumber\\
\delta \epsilon_y &=& \beta_y  \delta (y'^2)             \nonumber\\
\frac{d}{dt} <\epsilon_y> &=& \frac{\beta_y}{\gamma} \frac{d}{d\tilde{t}}
        <p_y^2/p_0^2>      \nonumber\\
\end{eqnarray}
In Eq.(32), $y'=p_y/p_0$, $\delta \epsilon_y$ is the change in $\epsilon_y$ in a 
scattering event. 

For the longitudinal emittance one finds
\begin{eqnarray}
\epsilon_s  &=& [s^2/\gamma^2+(\beta_s \gamma p_s/p_0)^2]/\beta_s   \nonumber\\
\delta \epsilon_s &=& \beta_s  \delta (\gamma p_s/p_0)^2 \nonumber\\
\frac{d}{dt} <\epsilon_s> &=& \beta_s \gamma \frac{d}{d\tilde{t}}
        <p_s^2/p_0^2>             \nonumber\\
\end{eqnarray}
In Eq.(33), $s,p_s$ are the coordinates in the rest system and I have used the relationship
$(p-p_0)_{LAB}=(\gamma p_s)_{REST}$

For the horizontal emittance one finds
\begin{eqnarray}
\epsilon_x  &=& \{ [x-\gamma D p_s/p_0]^2 +[\beta_x (p_x/p_0-\gamma D'p_s/p_0)+
     \alpha_x (x-\gamma D p_s/p_0)]^2 \} /\beta_x    \nonumber\\
   &=& \{ [x-\gamma D p_s/p_0]^2 +[\beta_x p_x/p_0+ \alpha_x x - 
          \bar{D}\gamma p_s/p_0]^2 \} /\beta_x    \nonumber\\
   &=& \{ x^2+(\gamma D p_s/p_0)^2-2x\gamma D p_s/p_0+(\beta_x p_x/p_0+ \alpha_x x)^2+
         (\bar{D}\gamma p_s/p_0)^2-    \nonumber\\
   & &   2(\beta_x p_x/p_0+\alpha_x x)(\bar{D}\gamma p_s/p_0)  \}   /\beta_x    \nonumber\\
\delta \epsilon_x &=& \delta \{\beta_x^2  (p_x/p_0)^2  + \gamma^2 (D^2+\bar{D}^2) (p_s/p_0)^2 
        -2 \beta_x \bar{D} \gamma (p_x/p_0)( p_s/p_0) \} / \beta_x    \nonumber\\
\frac{d}{dt} <\epsilon_x> &=& \frac{\beta_x}{\gamma} 
        \frac{d}{d\tilde{t}} <p_x^2/p_0^2> + \frac{D^2+\tilde{D}^2}{\beta_x} \gamma
         \frac{d}{d\tilde{t}}<p_s^2/p_0^2> 
        -2 \tilde{D} \frac{d}{d\tilde{t}}<p_x p_s/p_0^2>     \nonumber\\
\end{eqnarray}
In the result for $\delta \epsilon_x$, the terms that are linear in $p_x$ or $p_s$ have been dropped
as they do not contribute to $<\delta \epsilon_x>$ . In a scattering event involving two particles
, the $\delta p_x$ of one
particle is equal and  opposite to the $\delta p_x$ of the other particle. This is also true
for $p_s$.

\section*{Acknowledgements} 

I thank I. Ben-Zvi for his comments and encouragement. 
I also thank A. Fedotov and Y. Eidelman for information regarding their results.
 
\section*{References}

1. G.Parzen, BNL report C-A/AP/N0.150 (2004)

2. A. Piwinski Proc. 9th Int. Conf. on High Energy Accelerators (1974) 405,
   M. Martini CERN PS/84-9 (1984), A. Piwinski Cern 87-03 (1987) 402,
   A. Piwinski CERN 92-01 (1992) 226

3. J.D. Bjorken and S.K. Mtingwa, Part. Accel.13 (1983) 115, K. Kubo and
   K. Oide  Phys. Rev. S.T.A.B., 4, (2001) 124401  

\end{document}